\documentclass[12pt]{iopart}
\usepackage{graphicx}
\usepackage{epstopdf}
\usepackage{psfrag}
\usepackage{subfigure}
\begin{document}

\title{Non-equilibrium many-body effects in driven nonlinear resonator arrays}

\date{\today}

\author{T Grujic$^1$, S R Clark$^{2,1}$, D Jaksch$^{1,2}$ and D G Angelakis$^{2,3}$}
\address{$^1$ Clarendon Laboratory, University of Oxford, Parks Road, Oxford OX1 3PU, UK}
\address{$^2$ Centre for Quantum Technologies, National University of Singapore, 2 Science Drive 3, Singapore 117542}
\address{$^3$ Science Department, Technical University of Crete, Chania, Crete, Greece, 73100}
\ead{t.grujic1@physics.ox.ac.uk,dimitris.angelakis@gmail.com}

\begin{abstract}
We study the non-equilibrium behavior of optically driven dissipative coupled resonator arrays. Assuming each resonator is coupled with a two-level system via a Jaynes-Cummings interaction, we calculate the many-body steady state behavior of the system under coherent pumping and dissipation. We propose and analyze the many-body phases using experimentally accessible quantities such as the total excitation number, the emitted photon spectra and photon coherence functions for different parameter regimes. In parallel, we also compare and contrast the expected behavior of this system assuming the local nonlinearity in the cavities is generated by a generic Kerr effect rather than a Jaynes-Cummings interaction. We find that the behavior of the experimentally accessible observables produced by the two models differs for realistic regimes of interactions even when the corresponding nonlinearities are of similar strength. We analyze in detail the extra features available in the Jaynes-Cummings-Hubbard (JCH) model originating from the mixed nature of the excitations and investigate the regimes where the Kerr approximation would faithfully match the JCH physics. We find that the latter is true for values of the light-matter coupling and losses beyond the reach of current technology. Throughout the study we operate in the weak pumping, fully quantum mechanical regime where approaches such as mean field theory fail, and instead use a combination of quantum trajectories and the time evolving block decimation algorithm to compute the relevant steady state observables. In our study we have assumed small to medium size arrays (from 3 up to 16 sites) and values of the ratio of coupling to dissipation rate $g/\gamma \sim 20$  which makes our results implementable with current designs in Circuit QED and with near future photonic crystal set ups.
\end{abstract}

\section{Introduction}
\label{sec:introduction}
Coupled resonator arrays (CRAs) interacting with single two-level systems embedded in each resonator have recently emerged as an exciting new platform for the realization of novel quantum many-body effects. CRAs may offer several features complementary to those of the successful and well-established `toolbox' of cold atoms in optical lattices \cite{fisher1989boson,jaksch1998cold}, such as single-site addressibility and intrinsic non-equilibrium physics. This has led to proposals to realize a variety of phenomena of great interest in condensed matter physics such as Mott transitions \cite{angelakis2007photon,hartmann2006strongly,greentree2006quantum}, effective spin models \cite{kay2008reproducing} and fractional quantum Hall states \cite{cho2008fractional}, among others \cite{hartmann2008quantum,tomadin2010many}. Though promising efforts are underway to build the first CRA systems, a number of technical challenges originating  from the existence of strong dissipation must be overcome before equilibrium physics can be explored. On the other hand, the natural open nature of CRAs and the inherent ability to address and observe single resonators make this system an ideal test-bed to study many-body quantum lattice models out of equilibrium beyond the canonical Bose-Hubbard (BH) model and its realizations in optical lattices.
 
In this work we look for signatures of  underlying many-body phenomena in CRAs by analyzing observables measured in the system's non-equilibrium steady state (NESS). These include optically accessible observables like photon spectra and correlation functions. We focus on small to medium size CRAs of a few sites implementable with current or near future experimental technologies. We study coupled single mode resonators each interacting with a two-level system, a setup known now as the Jaynes-Cummings-Hubbard (JCH) model \cite{angelakis2007photon}. This simple model, in contrast to proposals involving multi-level atomic systems and external fields \cite{hartmann2006strongly}, may be realized in a variety of technologies ranging from quantum dots \cite{badolato2005deterministic,hennessy2007quantum} embedded in coupled defects in photonic crystals \cite{busch2007periodic, notomi2008large} to  coupled superconducting transmission line resonators \cite{wallraff2004strong} interacting with superconducting qubits \cite{blais2004cavity,astafiev2010resonance,frunzio2005fabrication}. In our study we have assumed small to medium size arrays (from 3 up to 16 sites) and values of the ratio of coupling to dissipation rate $g/\gamma \sim 20$  which  makes our results implementable with current designs in Circuit QED \cite{houck2012chip} and with near future set ups involving fiber coupled cavities \cite{lepert2011arrays} and photonic crystals \cite{notomi2010manipulating}. 

The JCH as implemented in CRAs (with only the photonic excitations allowed to hop between neighboring cavities) motivated parallels with the predictions of the BH model. The latter naturally emerges in CRAs when one assumes generic nonlinear resonator effects instead of a Jaynes-Cummings interaction \cite{carusotto2009fermionized,ssn2010signatures, hartmann2010polariton,gerace2009quantum}. Indeed, the JCH and BH share many similarities, both describing bosons hopping coherently between nearest neighbor sites, with local nonlinearities. An equilibrium quantum phase transition between Mott-insulating and superfluid-like phases exists for both models. However, the BH Hamiltonian involves only a single species of bosons (photons in this case), while the excitations of the JCH model have both photonic and atomic components. Consequently, the equilibrium physics of the JCH model is expected to be richer as shown in \cite{koch2009superfluid,rossini2007mott,mering2009analytic,quach2009band,makin2008quantum}. Studying the JCH out of equilibrium, as naturally implemented in open driven CRAs, is likely to highlight even more interesting differences with novel features beyond the realm of the driven BH model. In addition, most existing work to date in out of equilbrium CRAs has used mean-field theory to treat the system \cite{tomadin2009non, ssn2010signatures,liu2011quantum,diehl2008quantum,diehl2010dynamical,tomadin2011nonequilibrium}. Going beyond mean-field theory is crucial to provide physically accurate insights, especially in the experimentally realistic few-resonator regime. This requires a faithful representation of the full Liouville space of the resonator system which in most cases beyond two resonators becomes challenging. For similar reasons of complexity, existing works to date on non-equilibrium resonator arrays exploring correlations and anti-bunching effects were always limited to minimal systems of two resonators \cite{ferretti2010photon,liew2010single,bamba2011origin,angelakis2009steady,angelakis2010coherent}. More recent work includes a study of the fluorescence spectrum of again two coupled Jaynes-Cummings resonators \cite{knap2011emission,nissen2012non}. The possibility of simulating gauge fields in driven dissipative Jaynes-Cummings arrays was investigated in \cite{nunnenkamp2011synthetic}, and artificial gauge fields in multi-resonator arrays in Bragg reflector micro-cavities assuming a Kerr interaction have also been studied recently \cite{umucalılar2011artificial}. 

In the present work we simulate CRAs beyond the two resonator regime, always assuming the full JCH model, by exploiting a combination of the matrix product state time-evolving block decimation (TEBD) algorithm \cite{vidal2003efficient, vidal2004efficient} and quantum trajectories.  We propose and analyze the many-body phases using experimentally accessible quantities such as the total excitation number, the emitted photon spectra and photon coherence functions for different, experimentally feasible, parameter regimes.
In parallel, we also compare and contrast the expected behavior of this system when assuming that the local nonlinearity in the cavities is generated by a Kerr effect rather than a Jaynes-Cummings interaction. We find that the physics predicted by the two models differs for realistic regimes of interactions even if one meaningfully matches the respective strengths of the model nonlinearities. We analyze in detail the extra features of the JCH model originating from the mixed nature of the excitations and investigate the regimes where the Kerr description would faithfully match the JCH predictions. We find that the latter is only possible for values of the light-matter coupling and losses beyond the reach of current technology. 
 
The structure of this manuscript is as follows.  In the first section we  describe the system and the dissipation/driving mechanisms. In Section 2 we discuss the local eigenstates and outline a  scheme for mapping the effective nonlinearities between the models. In Section 3 we investigate the steady state spectra of single driven dissipative resonators, and demonstrate that Kerr  physics can be achieved as a limiting case of the Jaynes-Cummings model. Section 4 treats arrays of a few sites as could be implemented in near future experimental setups, and results for yet larger arrays are presented in Section 5. In Section~\ref{sec:novel_effects} we investigate the behavior of the system in the strongly repulsive regime and look for signatures of photon fermionization and crystallization in the JCH model.  We then conclude in Section~\ref{sec:discussion}. 

\section{The system}
\label{sec:models}
\begin{figure}
  \centering
\subfigure{\includegraphics[width=0.45\textwidth]{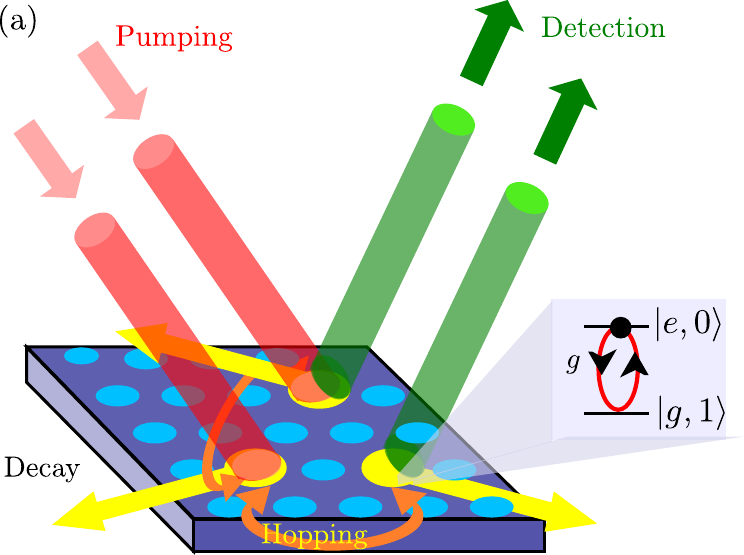}}
\hspace{0.05\textwidth}
\subfigure{\includegraphics{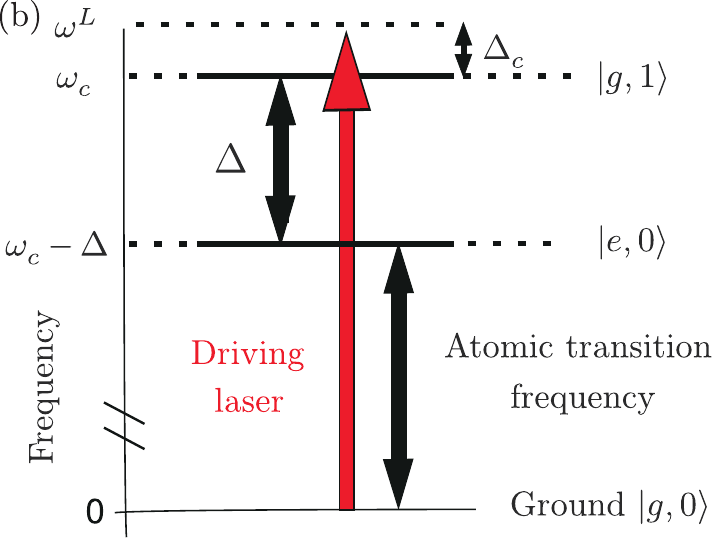}}\\
\subfigure{\includegraphics{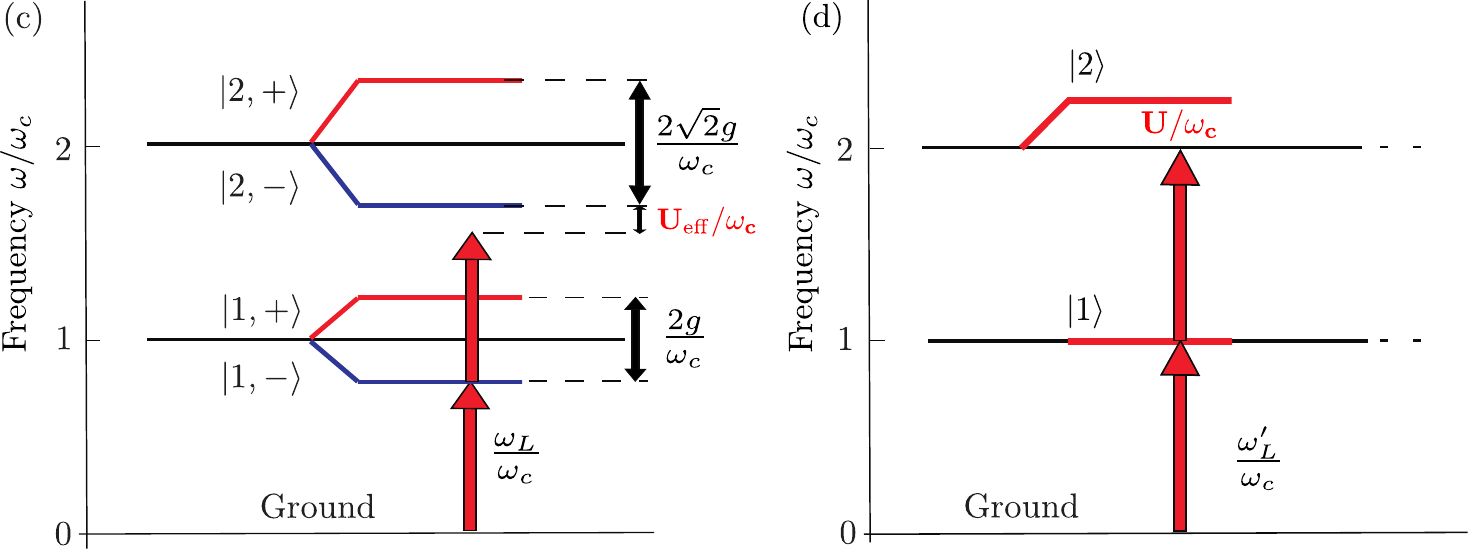}}
\caption{(a) A schematic of a three site set up. Here two-level systems are embedded in  cavities in a 2D photonic crystal but alternative implementations involving Circuit QED architectures could be also envisaged \cite{houck2012chip}. Coherent photon hopping between cavities results from modal overlap between neighboring cavities and photons can be lost due to decay mechanisms. (b) The relevant frequency scales and detunings used to describe a driven Jaynes-Cummings resonator. (c) The low-lying eigen-levels for the Jaynes-Cummings interaction, showing the definition of the effective Kerr nonlinearity $U_{\rm eff}$. (d) The low-lying excitations of a Kerr-nonlinear resonator. Arrows schematically show that a driving laser resonantly exciting a single-particle mode from the ground state is detuned from higher modes. Solid horizontal lines represent multiples of the cavity frequency. Dotted horizontal lines are a guide to the eye. }
\label{fig:system_schematics}
\end{figure}

We study $M$ coupled single-mode resonators (indexed by $j$) in a circular configuration, as shown schematically in Fig.~\ref{fig:system_schematics}~(a) for $M=3$.   The resonator frequency $\omega_c$ is the same for each resonator $j$. Coherent photon hopping between nearest-neighbor oscillators at a rate $J$ arises as a result of an overlap between neighboring resonator modes $\langle j, j' \rangle $. A large lattice spacing, relative to optical lattice setups, means that external driving lasers with amplitude $\Omega_j$ and frequency $\omega_L$ can independently excite and probe each resonator $j$. Assuming each resonator is coherently interacting with a two-level system, the physics is well described by the JCH model as defined below. We will analyze the non-equilibrium system response and in parallel contrast the results with the case where a generic Kerr nonlinearity is assumed in place of the Jaynes-Cummings interaction as described by the well known BH model. 

 The local physics for both cases is captured by the two Hamiltonians \footnote{ Assuming the spacing between neighboring modes within a resonator to be much larger than all other scales we only employ a single photon mode per resonator.}
\begin{eqnarray}
\hat{h}'_{JC} & = & \omega_c \hat{a}^\dag \hat{a} + \left ( \omega_c - \Delta \right ) \hat{\sigma}^+ \hat{\sigma}^- + g \left ( \hat{a}^\dag \hat{\sigma}^- + \hat{a} \hat{\sigma}^+ \right ),\\
\hat{h}'_{BH} & = & \omega_c \hat{a}^\dag \hat{a} + \frac{U}{2} \hat{a}^\dag \hat{a}^\dag \hat{a} \hat{a},
\end{eqnarray}
where we have set $\hbar=1$. The difference between the resonator frequency, and the atomic transition frequency is denoted by $\Delta$ and illustrated in Fig.~\ref{fig:system_schematics}~(b). The operator $\hat{a}^\dag$ creates a photon, while the operators $\hat{\sigma}^\pm$ denote the usual raising and lowering operations between the ground and excited states of the two-level system. 

The eigenstates of $\hat{h}'_{\rm JC}$ are known as `dressed' states, or polaritons. They are mixed atom-photon excitations, which are also eigenstates of the total excitation number operator $\hat{\mathcal{N}} = \hat{\sigma}^+ \hat{\sigma}^- + \hat{a}^\dag \hat{a}$ with eigenvalue $n$. The ground state is $|0\rangle \equiv |g,0\rangle$, with energy $E_0 = 0$. For a given excitation number $n \ne 0$, there are two polaritonic modes, designated $|n, \pm \rangle$, with associated frequencies in the bare frame $\omega_n^\pm = n \omega_c - (\Delta / 2) \pm \sqrt{\left ( \Delta / 2 \right )^2 + n g^2}$. The eigenstates can be written in the bare resonator-atom basis as $|n,\pm\rangle = \alpha_n^\pm |e,n-1\rangle + \beta_n^\pm |g,n\rangle$, with $\alpha_n^\pm = (\chi_n \mp \Delta) / (\sqrt{2}\sqrt{\chi_n^2 \mp \Delta \chi_n})$, and $\beta_n^\pm = \pm 2 g \sqrt{n}/(\sqrt{2}\sqrt{\chi_n^2 \mp \Delta \chi_n})$ \cite{barnett2003methods}. Here $\chi_n = \sqrt{\Delta^2 + 4 n g^2}$. It can be seen from these coefficients that setting a larger positive $\Delta$ results in an enhanced atomic component of the `-' polaritons, while negative $\Delta$ increases the photonic contribution, with the relative composition of polaritons reversed for the `+' species. The increased splitting of the polaritonic frequencies from the bare resonator frequency with increasing $n$ gives rise to an effective interaction between incoming photons, leading to a well known phenomenon in Cavity QED, the  photon blockade \cite{birnbaum2005photon}. Figure~\ref{fig:system_schematics}~(c) shows the eigen-structure schematically for the particular case $\Delta/g = 0$, illustrating the energy mismatch between adjacent polariton manifolds. The resonance condition for an $n$-photon excitation of the $|n, \pm \rangle$ polaritonic mode in a single resonator from the ground state $|0 \rangle$ is to set the driving laser detuning
 $\Delta_c = \omega_L - \omega_c = \omega_n^\pm / n - \omega_c$ at
\begin{equation}
\left ( \Delta_c \right )_n^\pm = \frac{1}{2 n} \left ( \Delta \pm \chi_n \right ).
\label{eq:driving_condition}
\end{equation}

The local Hamiltonian for the driven BH model, $\hat{h}_{\rm BH}'$, describes a single-mode resonator with a Kerr-type nonlinearity in the particle number of strength $U$. The eigen-frequencies are $\omega_n = n \omega_c + \frac{U}{2} n (n-1)$, with corresponding eigenstates being the Fock number states $|n \rangle$. The eigen-structure of the three lowest-lying levels is shown in Fig.~\ref{fig:system_schematics}~(d). An $n$-photon excitation of the $n^{\rm th}$ mode occurs for the laser detuning $\Delta_c =  \omega_L - \omega_c =  \omega_n / n - \omega_c$
\begin{equation}
(\Delta_c)_n = \frac{U}{2}(n-1).
\label{eq:driving_condition_BH}
\end{equation}

After transforming to a frame rotating at the driving laser frequency $\omega_L$ \cite{armen2006low}, the Hamiltonian for both model systems can be written generically as 
\begin{equation}
\hat{H}_X = \sum_{j=1}^M  \hat{h}_X^{(j)} + \sum_{j=1}^M \Omega_j \left ( \hat{a}_j^\dag + \hat{a}_j \right )  - J \sum_{<j,j'>} \hat{a}_j^\dag \hat{a}_{j'}.
\end{equation}
The first term describes the local physics in resonator $j$ in the rotating frame, with $X \in \{JCH, BH\}$. The second term describes the action of a coherent driving laser on each resonator $j$, and the last term describes the photon hopping. The local contributions $\hat{h}_{JCH}^{(j)},\hat{h}_{BH}^{(j)}$ are identical in form to $\hat{h}_{JC}',\hat{h}_{BH}'$, with the bare cavity frequency replaced by the detuning $\omega_c \rightarrow -\Delta_c$.

We now describe the manner in which we compare the two models. As the two Hamiltonians are patently different, care must be taken when comparing their behavior. We map the Jaynes-Cummings nonlinearity to an effective Kerr interaction via a frequency mismatch argument, as considered in both equilibrium and non-equilibrium contexts \cite{makin2008quantum,hãžmmer2012non}. In the following we focus on the `-' species of polaritons, as the driving laser frequency necessary to resonantly excite the $|n, - \rangle$ mode increases with $n$, qualitatively similar to a Kerr nonlinearity. We define the effective Kerr nonlinearity $U_{\rm eff} = U_{\rm eff}(g, \Delta)$ as the energy penalty incurred in forming a two-particle polaritonic excitation in a resonator (with energy $\omega_2^-$) from two one-particle polaritons in neighboring resonators (with total energy $2 \omega_1^-$) 
\begin{equation}
\frac{U_{\rm eff}}{g} = \frac{\omega_2^- - 2 \omega_1^-}{g} = \frac{\Delta}{2g} + 2 \sqrt{\left ( \frac{\Delta}{2g} \right )^2 + 1} - \sqrt{\left ( \frac{\Delta}{2g} \right )^2 + 2} .
\label{eq:eff_U}
\end{equation}
Figure~\ref{fig:system_schematics}~(c) shows these low-lying polaritonic eigenstates and the definition of $U_{\rm eff}$ pictorially, while Fig.~\ref{fig:system_schematics}~(d) shows the level structure for a Kerr-nonlinear resonator. Interestingly, an analogous definition for the `+' polaritonic branch yields an attractive Kerr-type interaction. We note that as the distribution of energy levels with particle number $n$ is markedly different between the models, it is only sensible to compare them in the very weakly excited regime where the above definition is meaningful. 

While the above mapping `matches' the nonlinearities of the two models by considering transitions between the one- and two-particle manifolds, we expect the single particle resonances of both systems to occur at different spectral locations. The dominant spectral features in weakly excited systems will therefore be observed for different driving laser parameter regimes. Additionally, the mapping involves a comparison only of diagonal elements of the two governing Hamiltonians. However the different natures of the excitations of the models are not fully accounted for in such a mapping, in particular the additional internal degree of freedom possessed by the JCH model. We therefore expect the corresponding distinct off-diagonal Hamiltonian terms to lead to different physical observables between the models even under this mapping.

We assume a finite photon loss rate $\gamma_p$ from each resonator for both models, and that the spontaneous emission rate from the excited level $|e \rangle$ of the two-level systems in the JCH system is negligible, $\gamma_a = 0$. Competition between coherent resonator driving, and photon loss leads to NESS conditions. We employ a master equation formalism \cite{carmichael1991open} to describe the evolution of the system's density matrix $\rho(t)$. The NESS density matrix $\rho_{\rm ss}$ is given by the stationary point of the master equation $\dot{\rho} = \mathcal{L}[\rho] = 0$, where the action of the Liouville super-operator is defined through
\begin{equation}
\mathcal{L}_X [\rho_{\rm ss}] = \frac{1}{i} [\hat{H}_X, \rho_{\rm ss}] + \sum_{j=1}^M \frac{\gamma_p}{2} \left ( 2 \hat{a}_j \rho_{\rm ss} \hat{a}^\dag_j - [\hat{a}^\dag_j \hat{a}_j, \rho_{\rm ss}]_+ \right ) = 0,
\label{eq:master_equation}
\end{equation}
with $[\cdot,\cdot]_+$ denoting the anti-commutator operation. In general, solving Eq.~(\ref{eq:master_equation}) is a formidable task, owing to the exponential growth in the necessary size of a system's description with the number of cavities $M$. We exploit the permutational symmetry of a homogeneous minimally sized three-site cyclic resonator system to enable solution via an exact diagonalization scheme, and employ a more sophisticated TEBD based stochastic unraveling of the master equation for larger systems. Details of the latter approach are provided in the Appendix.
\label{sec:single_spectra}
Before discussing the solutions of Eq.~(\ref{eq:master_equation}) for an array of cavities, we first revise and compare the stationary behavior of a single driven resonator with both Jaynes-Cummings and Kerr-type nonlinearities to illustrate their intrinsic differences in the local physics regime. 

\section{Local resonator physics: Kerr versus Jaynes-Cummings nonlinearities}
\label{sec:local}
\begin{figure}
  \centering
\subfigure{\includegraphics{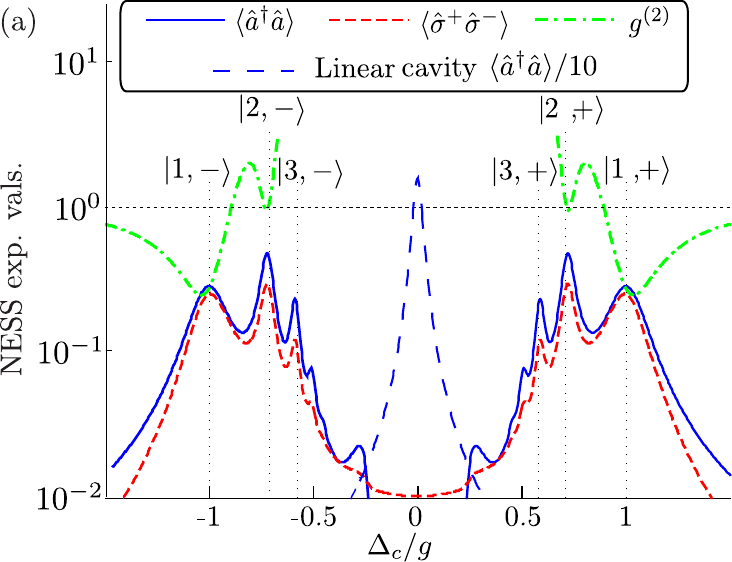}}
\subfigure{\includegraphics{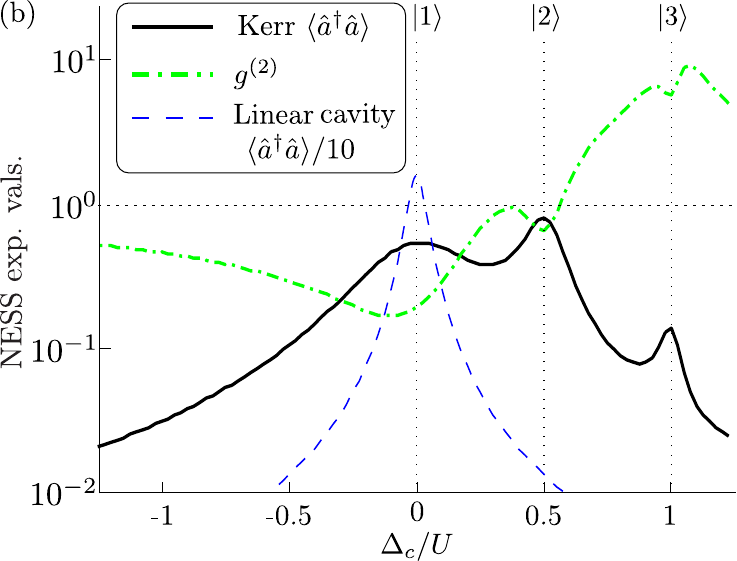}}\\
\subfigure{\includegraphics{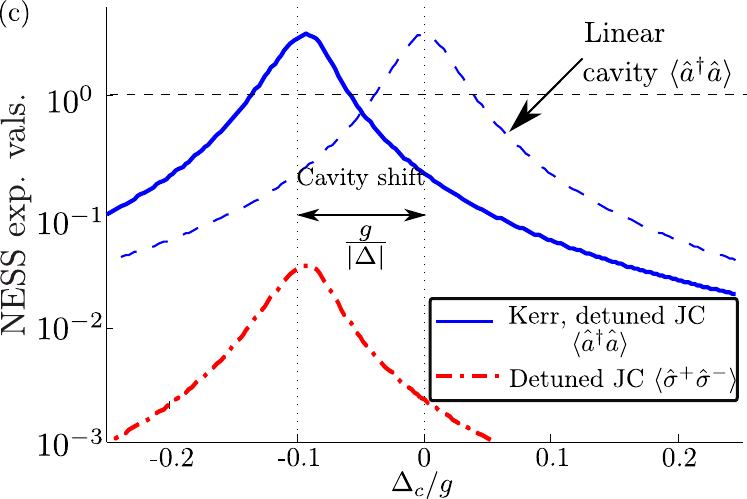}}
\subfigure{\includegraphics{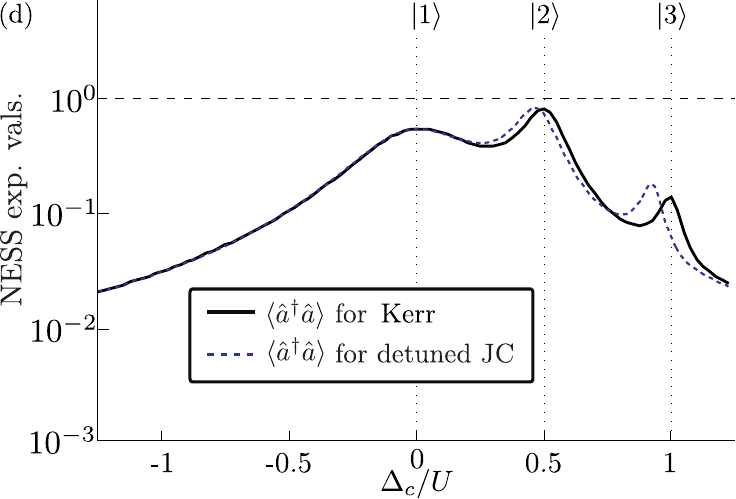}}
\caption{Steady state values of photonic/atomic population (solid/dashed line) and the second order photon correlation function ($g^{(2)}$, dot-dashed line) as a function of the driving laser-cavity detuning for a single driven dissipative resonator.  (a) Jaynes-Cummings resonator (atom resonant with the cavity mode $\Delta / g=0$). The central Lorentzian (dashed curve) is the response of an empty (linear) driven-dissipative resonator $\langle \hat{a}^\dag \hat{a} \rangle / 10$. (b) Kerr-nonlinear resonator. Vertical dashed lines and labels in both plots indicate resonator resonances according to Eqns.~(\ref{eq:driving_condition}) and (\ref{eq:driving_condition_BH}), respectively. Parameters: Both systems share $\Omega / \gamma_p= 2$. Jaynes-Cummings coupling $g / \gamma_p = 20$. For the Kerr resonator: $U / \gamma_p \approx 11.7$, set using Eq.~(\ref{eq:eff_U}). (c) The response of a far-detuned ($\Delta / g = -10$) Jaynes-Cummings resonator for identical atom-resonator coupling, driving and loss as above (d) Comparison of the Kerr response in (b) and that of a detuned Jaynes-Cummings resonator ($\Delta / g = -10$) with an ultra-strong nonlinearity $g' / \gamma_p \approx 1.6 \times 10^4$.}
\label{fig:example_spectra_single_cavity}
\end{figure}

In Figures~\ref{fig:example_spectra_single_cavity} (a) and (b) we show NESS particle numbers for both types of resonator as a function of the driving laser detuning from the bare cavity resonance. The most striking difference between the spectra is the existence of two `wings' for the Jaynes-Cummings nonlinearity, symmetrically distributed about the cavity frequency for this special case $\Delta / g = 0$. The driving laser frequency necessary to excite the `-' polaritons increases with the excitation number $n$ (the converse is true for the `+' polaritons). Figure ~\ref{fig:example_spectra_single_cavity} (b) also demonstrates that the Kerr resonator always exhibits a single particle response at the bare cavity frequency, whereas the spectral location of the two single particle modes in a Jaynes-Cummings type system strongly depend on the atom-resonator coupling parameters. 

The nonlinear response of the resonators can also be probed by analyzing the second order correlation function of the emitted photon from a certain resonator site. For later use, we define the generalized coherence function between any two resonators $(j,k)$ as $g^{(2)}(j,k) = \langle a_j^\dag a_k^\dag a_j a_k \rangle / \langle a^\dag_j a_j \rangle \langle a^\dag_k a_k \rangle$. In the following we refer to the on-site coherence function as $g^{(2)} \equiv g^{(2)}(j,j)$. As expected, in the strong coupling regime this quantity exhibits a dip below the coherent driving laser value $g^{(2)} = 1$   leading to photon anti-bunching when the driving laser is tuned to the corresponding single-particle modes for both models. For the Jaynes-Cummings resonator, the strongest anti-bunching is expected for laser detunings $\Delta_c = \pm g$ from the bare cavity frequency (for $\Delta = 0$). In contrast, the Kerr resonator demonstrates strongest anti-bunching when the driving laser is on resonance with the bare cavity mode, coinciding with the single particle mode at $\Delta_c = 0$. The reasonable atom-cavity coupling and cavity loss rates we have chosen give $g/\gamma=20$ (shown in Fig.~\ref{fig:example_spectra_single_cavity}), causing the correlation function to dip to a minimum $g^{(2)} \approx 0.25$ for the Jaynes-Cummings resonator, and to $g^{(2)} \approx 0.15$ for the Kerr resonator. 

Dips in the coherence function also appear when the driving laser is resonant with higher underlying quantum resonances (at equally spaced laser frequency intervals $U/2$ for the Kerr resonator, and laser frequencies $\Delta_c = \pm \frac{g}{\sqrt{n}}$ for the Jaynes-Cummings system). The magnitude of the correlation function at these resonances is highly sensitive to the magnitude of the driving laser strength and cavity loss rate.

Figures~\ref{fig:example_spectra_single_cavity}~(a) and (b) share several features common to driven dissipative nonlinear quantum systems. We see that, for both models spectral peaks corresponding to higher excitations $n$ become successively narrower, because the monochromatic driving laser cannot be simultaneously resonant with all intermediate levels. This means all but the lowest (polaritonic or Fock) peaks involve off-resonant transitions. The peaks also become less bright with increasing $n$ because in addition to being more difficult to populate, modes with more photons are more susceptible to photon loss. The Jaynes-Cummings system involves both atomic and photonic species, leading to different peak intensities between the models. This follows because we are only driving the photonic degree of freedom which in turn is coupled to the atomic degree of freedom in the Jaynes-Cummings resonator. 

Thus the spectral signatures of driven dissipative JCH and BH systems for the case of vanishing photon hopping (single resonators) are in general quite different. However, the two models do possess broad qualitative similarities in that they both describe the interplay of coherent bosonic hopping with an on-site nonlinearity. Indeed, the BH Hamiltonian has been widely used as an approximation to treat CRAs in several works assuming a generic nonlinearity  \cite{carusotto2009fermionized,ssn2010signatures, hartmann2010polariton,gerace2009quantum}. It is natural to ask if the two models are equivalent in some regime or even if the physics of the JCH Hamiltonian can be mapped on to an effective BH model. Such a mapping would enable the large body of existing knowledge about the BH model to be applied directly to CRAs and would additionally significantly simplify numerical simulation. As the nonlinearity acts locally inside each resonator, we address this question below by first investigating single resonators. 

\emph{The photonic limit of the Jaynes-Cummings model} ($\Delta \ll -g$) - 
Intuitively, we expect agreement between the models as the polaritonic JCH excitations are made more photonic in nature. For the `-' polariton species, this means setting $- g / \Delta\ll 1$. The two-level system in each resonator is then barely excited and the dispersive interaction available is capable of inducing a significant effective photon repulsion only for very large values of the coupling $g / \gamma_p \gg 1$. In this limit a Taylor expansion of the polaritonic eigen-energies $\omega_n^\pm$ in the small quantity $|g / \Delta|$ leads to frequencies quadratic in the particle number $n \ne 0$ as $\omega_{n,-}(g,\Delta) \approx n \left ( \omega_c  + g^2 / \Delta \right ) + g (g / |\Delta|)^3 n (n-1)$. We see that the off-resonant interaction induces an effective shift in the bare resonator frequency $\Delta_{\rm shift} = - g^2 / |\Delta|$, and that the energy spectrum can be written in the canonical Kerr nonlinear form with effective repulsion $U_{\rm BH}^{\rm approx} \equiv 2 g \left (g / |\Delta| \right )^3$. 

Figure~\ref{fig:example_spectra_single_cavity}~(c) shows the spectrum for a far detuned ($\Delta / g = -10$) driven dissipative Jaynes-Cummings resonator for the same experimentally realistic atom-cavity coupling rate $g / \gamma_p$ as in Fig.~\ref{fig:example_spectra_single_cavity}~(a). We see that the response is essentially that of an empty (linear) driven dissipative resonator, albeit with a peak response at $\Delta_{\rm shift}$ from the bare resonator frequency. The atomic excitation is an order of magnitude smaller, as expected. The effective Kerr nonlinearity is much smaller than the line-width ($U_{\rm BH}^{\rm approx} / \gamma_p = \frac{1}{50}$), so that nonlinear effects barely show up, apart from a slight asymmetry in the response. To observe sizable nonlinear effects for realistically achievable resonator parameters, we see that it is instead necessary to operate near the resonance point $\Delta / g \approx 0$. 

For comparison, the physically unrealistic ultra-strong atom-resonator coupling limit is shown in Fig.~\ref{fig:example_spectra_single_cavity}~(d). The response of the Kerr-nonlinear resonator shown in Fig.~\ref{fig:example_spectra_single_cavity}~(b) is reproduced, and compared with the spectrum of a Jaynes-Cummings resonator again operating in the photonic regime ($\Delta / g = -10$). The coupling strength $g / \gamma_p > 10^4$ is chosen by setting $U_{\rm BH}^{\rm approx} = U$ in the above definition of the effective Kerr nonlinearity. We see good agreement between the photon number spectra, which worsens for higher excitation peaks as higher order terms in the Taylor expansion of the polaritonic eigen-frequencies become important. So while deep in the regime $- g / \Delta \ll 1$ the Jaynes-Cummings model can, with reasonable accuracy, be mapped on to a Kerr system, this regime is inaccessible with current technology. 

\emph{The atomic limit of the Jaynes-Cummings model}  ($\Delta \gg g$) - 
In the opposite limit of $\Delta  / g \gg 1$, the atomic component of the `-' polaritons is maximal. A Taylor expansion of the eigen-frequencies now yields: $\omega_{n,-}(g,\Delta) \approx n \left ( \omega_c  - g^2 / \Delta \right ) + g (g / \Delta)^3 n (n-1) - \Delta$, for $n \ne 0$. The ground state $|g,0\rangle$ still lies at the zero of energy. Therefore we see that to first order in $g / \Delta$, the spectrum becomes equally spaced with the same shift in the resonator frequency as for the limit $\Delta \ll -g$, with the exception of the interval $\omega_c - \Delta$ between the ground state and the first polaritonic mode $|1,-\rangle$ . In this limit $|1,-\rangle \approx |e,0\rangle$, with a small photonic component allowing transitions by the external driving laser. This may be summarized by the following effective Hamiltonian describing the `vacuum shifted' $|n,-\rangle$ ladder of states for a single resonator in the atomic limit:
\begin{equation}
\hat{h}_{g / \Delta \ll 1} = (\omega_c - g^2 / \Delta) \hat{a}^\dag \hat{a} + g \left ( \frac{g}{\Delta} \right )^3 \hat{a}^\dag \hat{a}^\dag \hat{a} \hat{a} + \Delta |g,0 \rangle \langle g,0| - \Delta.
\end{equation}
Thus, population of the two-particle state from the resonantly driven single-particle mode is strongly inhibited. Higher lying levels could, however, be near resonantly populated from the one-particle mode given an additional driving laser with frequency $\omega_L = \omega_c$. 

When strictly working within the low-excitation regime, it is then appropriate to assign a large effective nonlinearity describing this energy penalty to reach the two-excitation manifold. However, care must be taken to distinguish between setups such as we consider, where only the lowest lying excitations are directly probed, and others which access the approximately harmonic ladder at larger $n>2$. 

\section{Many-body signatures in steady state observables}
\label{sec:finite_specta}
Moving beyond the single resonator regime we will first analyze a minimal cyclic nonlinear CRA of $M=3$ cavities. This case, though not truly many-body is interesting as this is where the first experimental implementations are likely to begin \cite{houck2012chip, lepert2011arrays, notomi2010manipulating}. We consider a finite coherent photon tunneling $J \ne 0$ between adjacent cavities which splits the local polaritonic resonances into delocalized global modes. For the moment we drive our system homogeneously so that all external lasers are in-phase with $\Omega_j = \Omega, \forall j$.
 
\begin{figure}
  \centering
\subfigure{\includegraphics[width=0.47\textwidth]{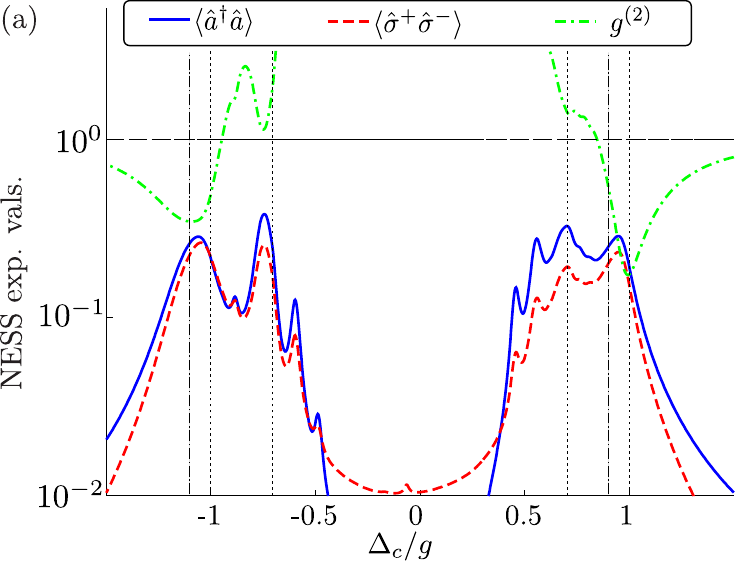}}
\subfigure{\includegraphics[width=0.47\textwidth]{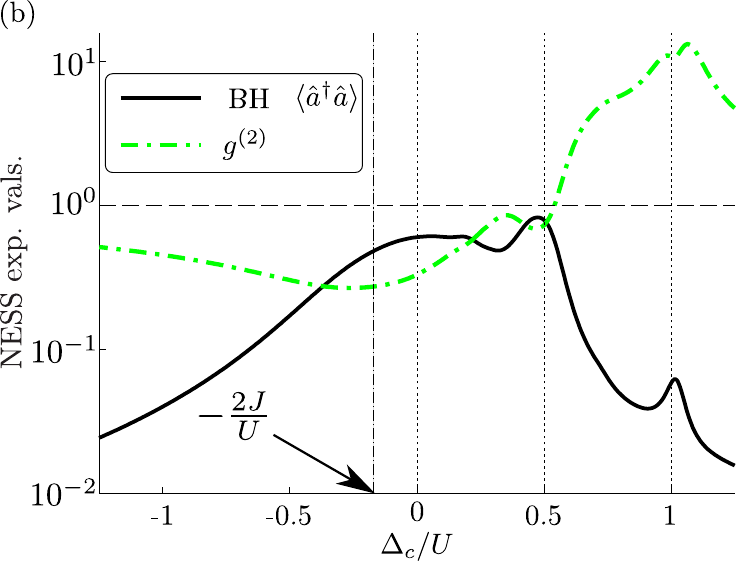}}\\
\subfigure{\includegraphics{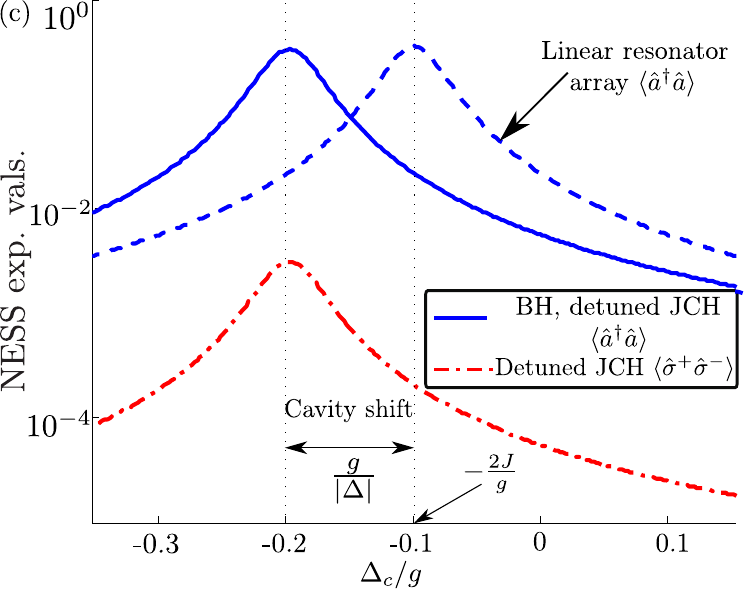}}
\subfigure{\includegraphics{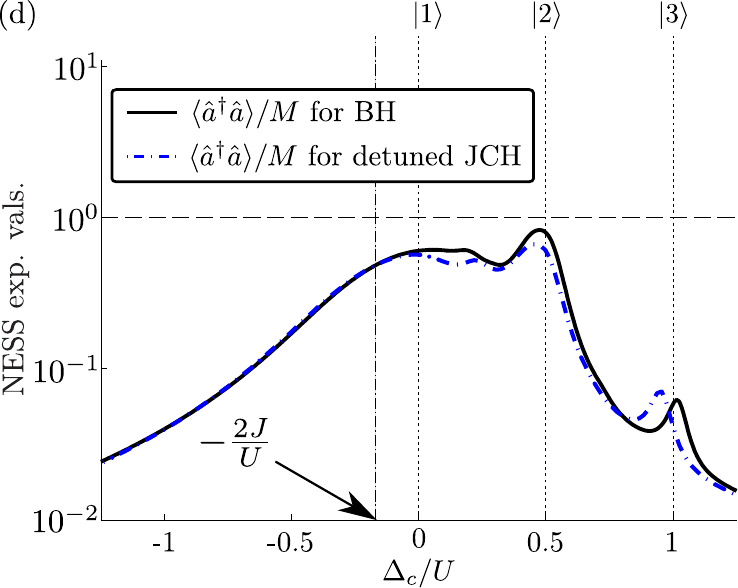}}
\caption{The steady state per-resonator photonic/atomic populations and $g^{(2)}$ function for homogeneously driven, cyclic three-site resonator systems described by (a) the JCH Hamiltonian (with $\Delta / \gamma_p= 2 J/\gamma_p = 2$) and (b) the BH model. Vertical dashed lines indicate the location of underlying modes of an isolated resonator ($J=0$), while vertical dash-dotted lines indicate the positions of delocalized single particle modes. The atom-resonator coupling $g / \gamma_p = 20$ for (a) and (b) is again as in Figs.~\ref{fig:example_spectra_single_cavity}, but there is now a finite photon hopping $J/\gamma_p = 1$. (c) and (d) are generalizations of the single resonator `photonic limit' spectra of Figs.~\ref{fig:example_spectra_single_cavity}, extended to three-resonator systems with the same photon hopping. The driving rate in (c) is set at $\Omega / \gamma_p = 0.3$ to enable numerical solution, otherwise all parameters are unchanged. Note in (d) the almost complete overlap of the response of the two many-body models now in this `photonic' limit. This agreement, however, requires an unrealistic ultra-strong value of $g / \gamma_p \approx 1.6 \times 10^4$. }
\label{fig:example_spectra_3_cavities}
\end{figure}

We obtain the NESS by diagonalizing the super-operator $\mathcal{L}$ after exploiting the permutational symmetry of the system to significantly decrease the number of unique density matrix elements and additionally retaingin only basis states allowing a maximum of $P = 4$ excitations in the system. \footnote{For linear cavities this permits the numerical analysis of driving strengths up to $\Omega / \gamma_p \approx 0.3$ without significant truncation errors. With a resonator nonlinearity $g, U \ne 0$, occupation of higher-lying levels is suppressed enabling accurate exact diagonalization results for yet stronger driving, confirmed by extensive quantum trajectory calculations as detailed in the Appendix.}

 Figures~\ref{fig:example_spectra_3_cavities} (a) and (b) show generalizations of the single resonator spectra discussed above for the relatively small rate of photon hopping $J / \gamma_p = 1$. We expect the response of the systems to strongly resemble the `local' physics in this regime, as the photon hopping is essentially a weak perturbation on top of the atom-cavity coupling.

The homogeneous nature of our chosen driving means that the total momentum of excitations in the NESS must be zero. The one-particle excitation in the BH system is then just the zero-momentum free-particle Bloch mode $|k \!\!= \!\!0 \rangle = \frac{1}{\sqrt{M}} \sum_{j=1}^M \hat{a}^\dag_j |0 \rangle$ excited at a laser detuning $(\Delta_c)_{|k=0>} = -2 J$, as the Kerr nonlinearity does not influence a single particle. In contrast a single excitation in the JCH system can be shared between atomic and photonic degrees of freedom leading to two delocalized generalizations of the $|1, \pm \rangle$ polaritons, denoted $|k \!= \!0, \pm \rangle = A^\pm |E\rangle + B^\pm |k \!\!= \!\!0 \rangle$. Here $|E \rangle = \frac{1}{\sqrt{M}}\sum_{j=1}^M \hat{\sigma}^+_j |0 \rangle$ is a delocalized atomic excitation. The coefficients $A^\pm$ and $B^\pm$ are identical in form to the coefficients $\alpha_1^\pm$, $\beta_1^\pm$ of the localized polaritons, after making the replacement $\Delta \rightarrow \Delta_{k=0} = \Delta - 2 J$. These are resonantly excited at laser detunings  $(\Delta_c)_{|k=0, \pm \rangle} = -2 J - \Delta_{k=0}/2 \pm \chi_{k=0}^{(1)}$, where $\chi_{k=0}^{(1)} = \sqrt{g^2 + \Delta_{k=0}^2/4}$. As only the cavities are directly coupled there is now an additional asymmetry between photons and atoms. The resonance point for delocalized polaritons is shifted to $\Delta = 2J$ from $\Delta=0$ found in the single resonator case. This is equivalent to setting the transition frequency of the atoms equal to the frequency of the lowest lying Bloch mode $|k \!\!= \!\!0 \rangle$. 

Relative to Figs.~\ref{fig:example_spectra_single_cavity}~(a) and (b), we see new multi-particle modes appear between the delocalized generalizations of the dressed state resonances. By an $N$-particle mode we mean an eigenstate of the total excitation number operator for the whole system $\hat{\mathcal{N}} = \sum_{j=1}^M \hat{\mathcal{N}}_j$ with eigenvalue $N$. For this particularly small photon hopping rate, equal to the cavity line-width, the delocalized single particle mode of the BH system is smeared into a broad hump along with delocalized two- and three- particle modes, while new features can also be discerned in the JCH spectra. Additionally, new modes appear in the JCH model at approximately the bare cavity frequency. These modes are symmetrized superpositions of `+' and `-' polaritons, and for our chosen driving strength are barely populated relative to the  response of the BH model at the bare cavity frequency. 

There is an asymmetry between the delocalized generalizations of the `-' and `+' wings of the JCH resonator array spectrum, despite setting $\Delta = 2J$. We note that for large photon hopping $J \gg g$ and $\Delta = 2J$, the resonator array spectrum is again symmetric about the Bloch mode frequency (not shown), though still with quantitative differences to the single resonator case. It is in the intermediate regime ($J \approx g$) that asymmetries appear, as hopping brings multiple particle global excitations into the same spectral region. A signature of this hopping-induced behavior that is particularly amenable to experimental verification is a measurement of the photon correlation at the two delocalized single-particle resonance frequencies. For this particular driving and cavity loss rate, the values are $g^{(2)} \approx 0.18, 0.35$ for the $|k=0, \pm \rangle$ modes respectively. Meanwhile $g^{(2)}$ reaches a minimum of $\approx 0.26$ at the underlying single-particle mode of the BH system (indicated by the vertical dash-dotted line in Fig.~\ref{fig:example_spectra_3_cavities}~(b)). We note here again that if one assumes losses smaller than $g/\gamma_p=20$ used here then the correlation minimum approaches zero and clear anti-bunching should be achieved, as  expected.

The spectral response of far-detuned finite-size CRAs operating in the photonic limit ($-g / \Delta \ll 1$) is shown in Figs.~\ref{fig:example_spectra_3_cavities} (c) and (d). These spectra are direct generalizations of those shown in Fig.~\ref{fig:example_spectra_single_cavity}. Again, we see that for our assumed atom-resonator couplings (i.e. $g / \gamma_p \approx 20$), the main effect of the atomic degree of freedom is to shift the free-particle mode by $\Delta_{\rm shift}$, in this case from the lowest Bloch mode. Figs.~\ref{fig:example_spectra_3_cavities} (d) shows that (unrealistically) larger couplings `matched' to a specific Kerr nonlinearity can reproduce the spectral features of a driven dissipative finite CRA with reasonable accuracy. 

\section{Larger resonator arrays} 
\begin{figure}
  \centering
\includegraphics{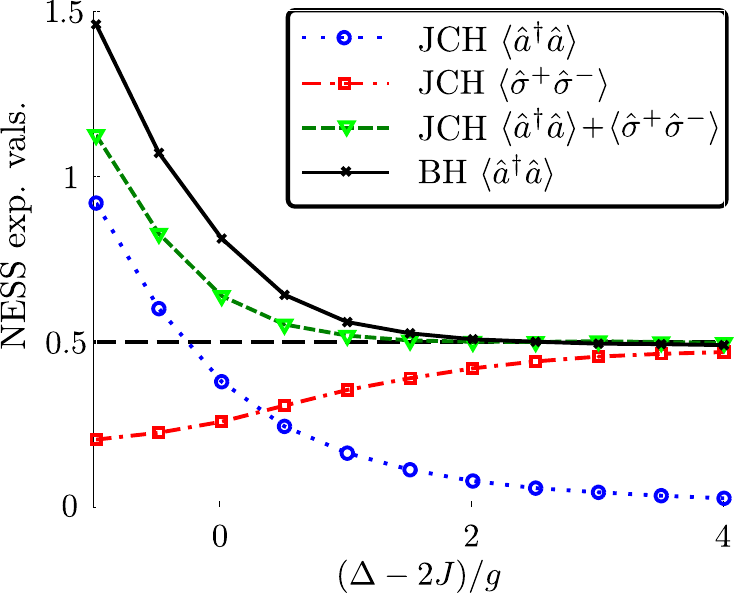}
\caption{NESS particle numbers per resonator for sixteen-site Bose-Hubbard and Jaynes-Cummings CRAs driven at their single particle resonances. Parameters: photonic driving $\Omega / \gamma_p = 2$, atom-resonator coupling strength $g / \gamma_p  = 20$, photon hopping rate $J / \gamma_p = 1$. Trajectory calculations retain $p = 6$ photons per resonator for the BH simulation, and $p=5$ for the JCH. Each Jaynes-Cummings detuning $\Delta / g$ is mapped to an effective Kerr nonlinearity via Eq.~(\ref{eq:eff_U}) to enable a comparison. }
\label{fig:8_site_system_NESS_expvals}
\end{figure}

So far we have investigated few sites JCH systems of a few sites operating either on-resonance (i.e. $\Delta / g = 0$), or in the photonic limit where the weak nonlinearity is captured by an effective BH interaction. We now demonstrate the effect of the changing nature of the polaritonic excitations and also investigate larger arrays of $M=16$ sites as this parameter is varied about the strong-interaction regime $\Delta / g \approx 0$. Rather than construct a spectrum by varying the driving laser frequency, we instead selectively drive a particular spectral feature and vary $\Delta / g$. Specifically, in Fig.~\ref{fig:8_site_system_NESS_expvals} we show trajectory results for NESS particle numbers for larger ($M=16$) JCH and BH arrays driven at their single particle resonances, corresponding to the driving laser detunings $\Delta_c = (\Delta_c)_{|k=0,-\rangle}$ and $\Delta_c = -2J$ respectively. 

As expected, we see the changing nature of the system's excitations reflected in a larger atomic (and smaller photonic) occupation for increasing $\Delta / g$. For strong positive detunings, the excitations are predominantly atomic, indirectly excited via the resonator field. As we have assumed lossless atoms, the steady state corresponds to oscillations between the ground and excited states, leading to an average half atomic occupancy. Under the mapping of Eq.~(\ref{eq:eff_U}), this regime corresponds to a BH system with large Kerr coefficient. Under the stronger driving we have chosen for this calculation, the BH system approximately oscillates coherently between zero and one photons per resonator, with occupation of higher Fock levels suppressed. Thus, while the total excitation number for the JCH asymptotically agrees with the NESS photon number for the BH in this limit, the underlying physics is very different. 

In the opposite limit $\Delta / g < 0$, the excitations of the JCH model become more photonic, and the effective Kerr nonlinearity decreases. We have already seen that agreement between the JCH and Kerr photon number is reached far into this limit. In the intermediate regime, $\Delta / g \approx 0$, the total particle number for the JCH exhibits significant departures from the BH results. This is due both to the fact that only one component of the polaritons is driven in the JCH system, and also that this stronger driving allows access to higher-lying states ($n>2$) outside the regime of validity for the effective Kerr strength definition. We therefore see non-trivial differences between the models both in spectra measured as a function of driving laser frequency, and as a function of the nonlinearity.

\section{The strong nonlinearity limit: fermionization and crystallization of photons in CRAs}
\label{sec:novel_effects}
Having demonstrated that the NESS in resonator arrays governed by the JCH and BH Hamiltonians are in general different we now evaluate two non-equilibrium effects in the large nonlinearity regime recently studied assuming a BH description, with proposals to experimentally realise such effects in Jaynes-Cummings type CRA systems.

\emph{Fermionization in coupled resonator arrays - } Recently an exploration of the ultra-strong nonlinearity regime in a three-site driven dissipative Bose-Hubbard model was undertaken \cite{carusotto2009fermionized}. For a closed coherent system it was found that as the Kerr strength $U$ approached infinity double occupancy of any resonator was completely suppressed allowing the system's bosonic wave-functions to be mapped to those of an equivalent fermionic system via a Jordan-Wigner transformation. In this interpretation double occupancy is prevented by fermionic statistics rather than a hard core bosonic interaction. The authors found that a cyclic system's $N$-particle eigenfunctions and corresponding energies can then be identified uniquely for a given total particle momentum. Subjecting the system to coherent driving and a finite particle loss rate from each resonator results in readily classifiable peaks in the resonator occupancy as a function of the driving laser frequency. 

\begin{figure}[t] 
  \centering
\subfigure{\includegraphics{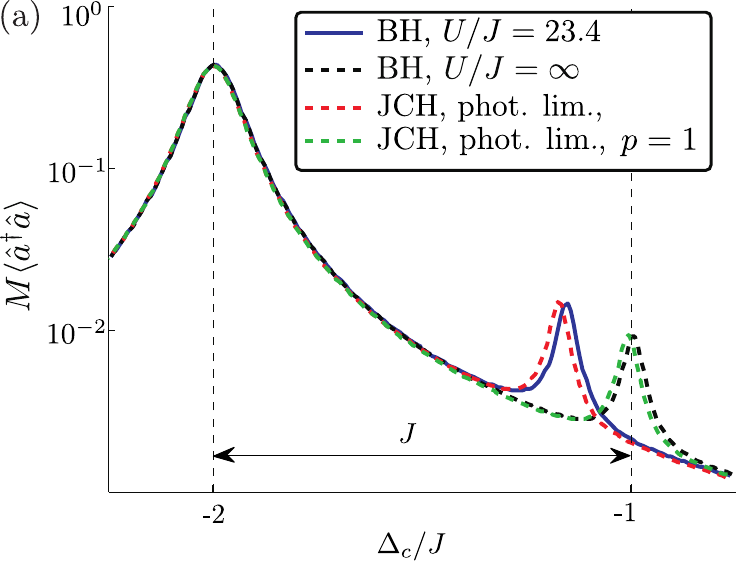}}
\subfigure{\includegraphics{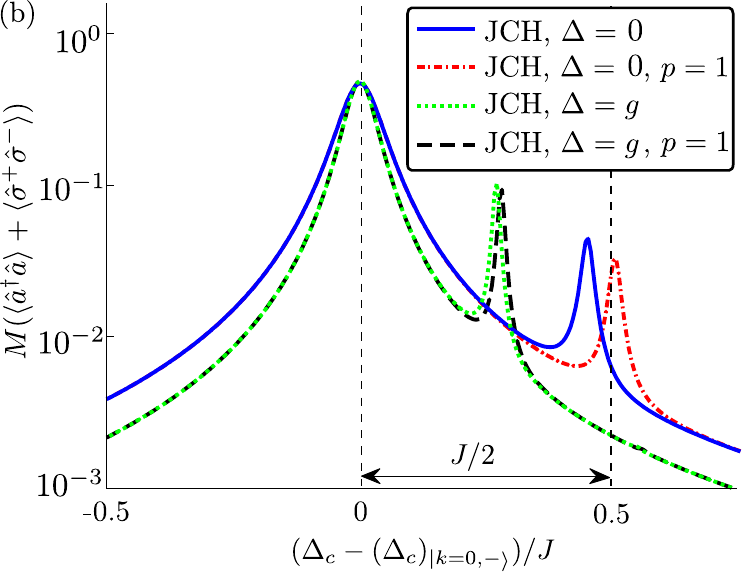}}\\
\subfigure{\includegraphics{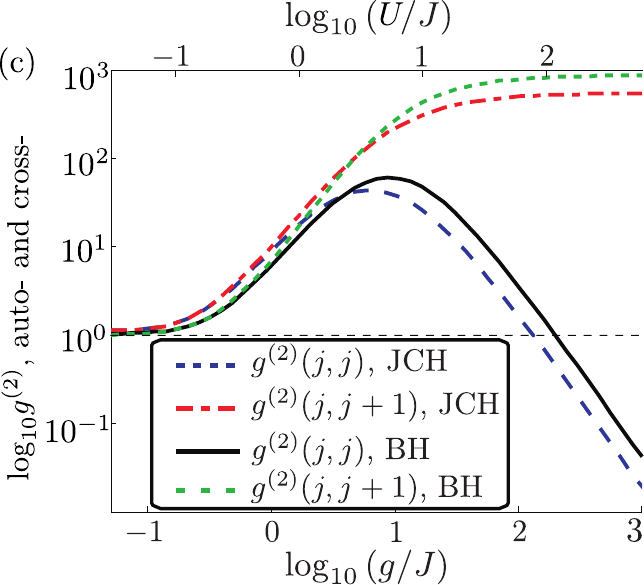}}
\caption{(a) The one and two particle particle peaks (right and left features, respectively) for a homogeneously driven three-resonator BH model in the strongly nonlinear regime (solid blue, $U / J = 23.4$), and the hard core limit ($p=1$, dotted black). Spectra for larger $U$ approach the hard core limit asymptotically. Parameters: photon hopping rate $J / \gamma_p = 20$, driving $\Omega / \gamma = 0.5$. Also shown are results for a strongly detuned JCH system operating in the photonic limit with $\Delta / g = -10$, and an unrealistically large atom-resonator coupling  $g/\gamma_p \approx 2 \times 10^5$. (b) Analogous photon number spectra for the JCH model with $g / \gamma_p \approx 800$ (chosen according to Eq.~\ref{eq:eff_U}), and the photon driving and tunneling rates as in (a), for two different atom-resonator detunings $\Delta / g$ (dashed and dash-dotted). Again, the hard-core spectra are recovered in the limit $g \rightarrow \infty$. The horizontal axis is taken relative to the single particle spectral position $(\Delta_c)_{|k=0,-\rangle}$, different for each $\Delta$. (c) The auto- and cross- photon correlation functions measured at the two particle peak as a function of the nonlinearity in both the BH and JCH ($\Delta = 2J$) Hamiltonians. JCH systems with different couplings $g$ are compared with Bose-Hubbard systems whose strength $U = U(g)$ (see upper horizontal axis) is obtained from Eq.~\ref{eq:eff_U}. Note these calculations are performed at the lower driving $\Omega / \gamma_p = 0.25$, to avoid truncation errors at low nonlinearity. }
\label{fig:comparison_of_BH_JCH_fermionised}
\end{figure}

Figure.~\ref{fig:comparison_of_BH_JCH_fermionised}~(a) reproduces such a spectrum for a homogeneously driven ($\Omega_j = \Omega,  \forall j$) minimal $M=3$ resonator BH system. For the spectral range shown the only possible modes that can be observed in the hard core limit NESS are the zero momentum one-particle mode, which for the BH model coincides with the Bloch mode $|k \!\!= \!\!0 \rangle$ of an empty resonator system, and a mode formed from two particles with opposite momenta $k = \pm 2 \pi / 3$. We show spectra for a strong nonlinearity $U / \gamma_p \gg 1$ (allowing for any number of photons per resonator in the calculation) and for the true fermionized limit $U \rightarrow \infty$ (when strictly $p=1$ photon per resonator is retained in calculations) illustrating the convergence. Also shown in Fig.~\ref{fig:comparison_of_BH_JCH_fermionised}~(a) are NESS photon numbers for a strongly detuned JCH system operating in the `photonic regime', again computationally retaining multiple and single photons per resonator. As expected from the discussion in Sec.~\ref{sec:local} good agreement with the BH results is observed, albeit for unrealistically large values of the light-matter coupling $g$. 

Moving beyond from the photonic limit we now explore the novel physics of ultra-strong nonlinearity in JCH systems operating in the regime of strongest interaction between resonators and atoms, i.e. $\Delta / g \approx 0$. In this regime correspondence between the models is less clear and the fully polaritonic nature of excitations must be taken into account. To recover physics which resembles hard-core photons we operate in the limit $g / \gamma_p \gg 1$, along with $g \ll \omega_c$ to satisfy the rotating wave approximation, and drive at $\omega_L \approx \omega_1^-$ near the $|1,-\rangle$ polaritonic resonance of isolated resonators. Under these conditions the only relevant states in each resonator are the ground state and $|1,-\rangle$ polaritonic state.

Figure~\ref{fig:comparison_of_BH_JCH_fermionised}~(b) shows hard-core JCH model spectra for two different atom-resonator detunings $\Delta = 0$ and $\Delta = g$. Increasing the coupling $g$ while holding the ratio $\Delta/ g$ fixed leads to a qualitatively similar evolution of spectral features as in the BH model. A two-particle excitation splits from the single particle resonance and asymptotically approaches a splitting that only depends on $J$ and $\Delta / g$. For the coupling $g/\gamma_p \sim 800$ used good agreement is seen in Fig.~\ref{fig:comparison_of_BH_JCH_fermionised}~(b) between the full JCH model and the spectrum obtained when retaining only the $p=1$ photons per resonator necessary to describe the hard-core polariton limit. We see that even in the ultra-strong non-linearity limit of the JCH model the internal structure of polaritons (governed by the parameter $\Delta / g$) still plays a crucial role in determining the location and amplitude of spectral features. This behavior is a result of the effective hopping, driving and loss rates for polaritons being different from the bare photon parameters. Indeed the effective polaritonic hopping rate is $J_{\rm pol} = |\beta_1^-|^2 J$, the effective driving rate is $\Omega_{\rm pol} = (\beta_1^-) \Omega$, while the loss rate is $\gamma_{\rm pol} = |\beta_1^-|^2 \gamma_p$, where $\beta_1^-$ is the coefficient controlling the photonic component of the $|1,-\rangle$ polaritons defined in Sec.~\ref{sec:models}. This reflects that the hopping, driving and loss processes involve only the photonic component of the polaritons. Thus, on resonance ($\Delta / g = 0 \Rightarrow \beta_1^- = \frac{1}{\sqrt{2}}$) we see that the peak separation is $J_{\rm pol} = J/2$  as seen in Fig.~\ref{fig:comparison_of_BH_JCH_fermionised}~(b). 

Focusing on experimentally measurable photonic quantities, in Fig.~\ref{fig:comparison_of_BH_JCH_fermionised}~(c) we track for both models as a function of their nonlinearity the photon density-density correlations measured at the two-particle peak on a single site via $g^{(2)}(j,j)$ and between neighboring sites via $g^{(2)}(j,j+1)$. The spectral location of the zero-momentum two-particle modes for the JCH and BH systems, are found using the results in Refs.~\cite{wong2011two} and \cite{javanainen2010dimer} respectively. As outlined in Ref.~\cite{carusotto2009fermionized} at small nonlinearities the two-particle peak resides within the one-particle spectral feature and so correlations inherit Poissonian statistics from the driving laser. Larger nonlinearities split the two-particle resonance from the one-particle peak, as shown in Figs.~\ref{fig:comparison_of_BH_JCH_fermionised}~(a) and (b), and lead to strong bunching. For very strong nonlinearities on-site anti-bunching is expected as the two excitations are distributed in such a way that the pair of photons are never in the same resonator resulting in the nearest neighbor correlations becoming large. Such considerations, being consequences of generic nonlinear behavior, lead to qualitatively similar correlation functions for both the driven dissipative JCH and BH models. Thus fermionized photons are a feature of the on-resonance JCH model as well, once the different spectral frequencies for correlation measurements are taken into account. 

\emph{Polariton crystallization -} 
Another intriguing phenomenon of interacting photons in resonator systems, photon crystallization, was recently predicted to occur in a one dimensional ring of optical cavities with Kerr-type nonlinearity \cite{hartmann2010polariton}. Driving lasers with a phase difference of $\pi / 2$ between each site $k$, i.e. $\Omega_k = \Omega \exp (i k \pi / 2)$, create a flow of bosons around the system. The contact interaction energy $U$ was found to result in a `crystallization' of particles as they flowed around the ring, even in the presence of dissipation. The signature of this effect was identified in the particle density-density correlations $g^{(2)}(j,k)$ between cavities $j$ and $k$, measured in the system's steady state. 

On-site anti-bunching is accompanied by nearest-neighbor density-density correlations stronger than correlations between more distant cavities. The conclusion drawn in Ref.~\cite{hartmann2010polariton} was that particles form `dimers' of light, which flow around the system. We demonstrate here that this dimerization can also be seen in a JCH system outside the photonic limit. The additional degree of freedom in the atom-resonator detuning $\Delta$ allows the strength of this effect to be adjusted on demand. 

We study a system of $M = 16$ coupled Jaynes-Cummings resonators under periodic boundary conditions. In addition to the relative phases, we must choose the frequency of our driving lasers. For a BH-type system, a laser detuning $\Delta_c = 0$ directly drives the single particle $k = \pi / 2$ mode. For the JCH model we choose to drive the delocalized generalization of the single-particle polaritonic mode $|1, - \rangle_{k=\pi/2}$ to produce an analogous effect. Details of the parameters used for this numerically demanding calculation are given in the Appendix. 

\begin{figure}[t]
  \centering
\subfigure{\includegraphics{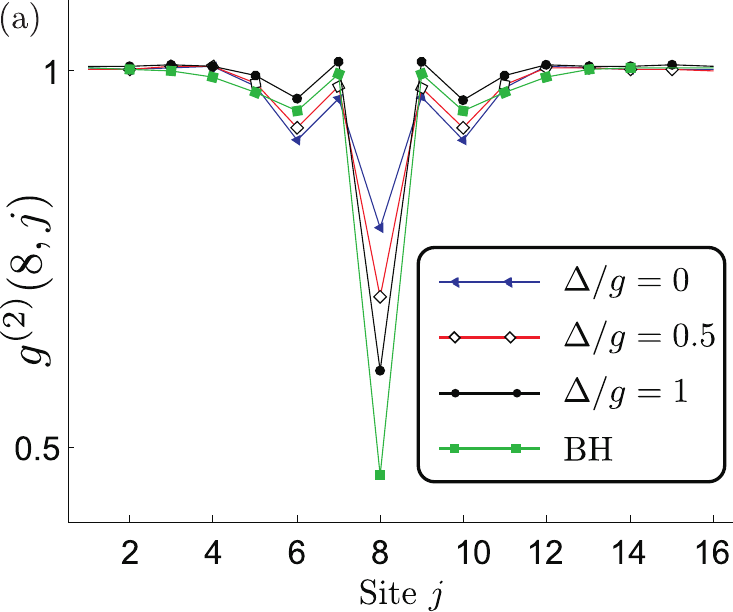}}
\subfigure{\includegraphics{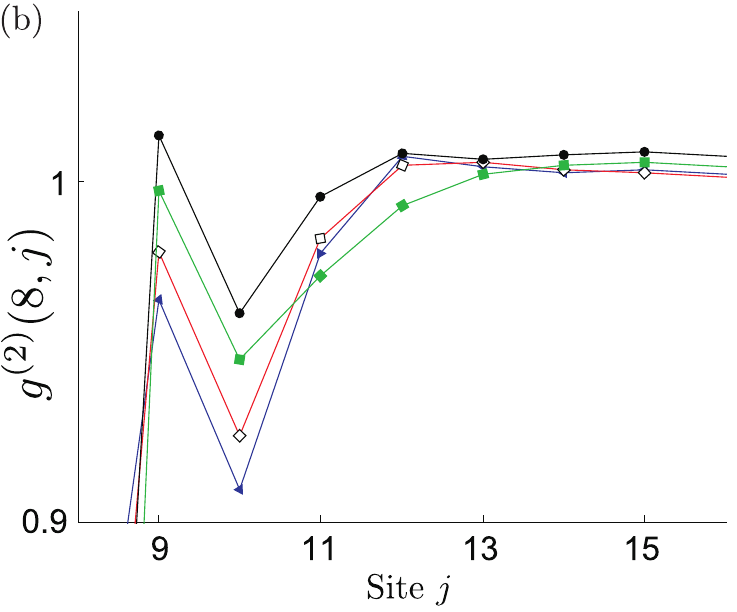}}\\
\subfigure{\includegraphics{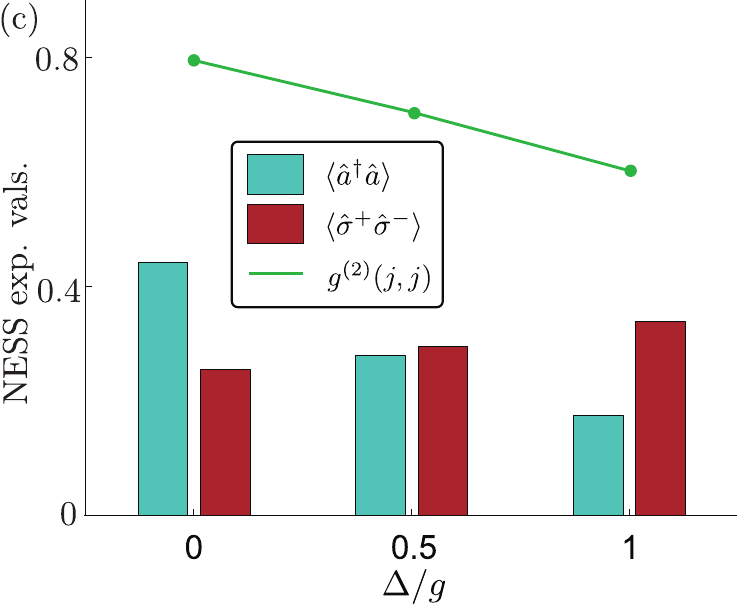}}
\caption{(a) Steady state photon density-density correlations for a cyclic 16 site JCH system, driven by lasers exciting the $\pi / 2$ momentum mode of the system. Thin lines between markers are drawn to guide the eye. Parameters: hopping $J / \gamma_p = 2$, atom-resonator coupling $g / \gamma_p = 10$, driving strength $|\Omega_k| / \gamma_p = 2$. Also shown (solid green) are density-density correlations for a $M=16$ site system with `matched' Kerr nonlinearity $U/\gamma_p \approx 6$ in each resonator, and otherwise identical parameters. (b) Closer view of the correlation functions. (c) The relative atomic and photonic population in each resonator in the steady state for the three different values of atom-resonator detuning chosen, as well as the on-site photon correlation function. }
\label{fig:polariton_crystallization}
\end{figure}

We see in Fig.~\ref{fig:polariton_crystallization}~(a) that the signatures of crystallization show up in our polaritonic system -- there is a larger probability of finding photons in neighboring cavities than in cavities further apart. We have additional control over the atom-resonator detuning $\Delta$, which allows us to tune the nature of the system's excitations either more photonic $\Delta / g < 0$, or atomic $\Delta / g > 0$. We see that the crystallization effect becomes weaker for larger $\Delta$ as the atoms and resonators are tuned away from resonance, while on-site anti-bunching is enhanced. This is a consequence of resonator photons being strongly coupled to the two-level systems, which hold most of the excitation for large $\Delta$, but longer-range correlations tend to unity. Figure~\ref{fig:polariton_crystallization}~(b) shows the actual photonic and atomic expectation values per resonator, for the three detunings considered illustrating the changing make-up and associated enhanced photon anti-bunching. Also shown for comparison in Figure~\ref{fig:polariton_crystallization} are results for the driven dissipative Kerr system like that considered in Ref.~\cite{hartmann2010polariton}. We conclude that the `dimerization' predicted there for the BH model persists in CRA calculations using the full atom-resonator Hamiltonian without approximations, and that some degree of tunability should be observable by bringing the cavities in and out of resonance while maintaining significant resonator populations. 

\section{Discussion}
\label{sec:discussion}
We have proposed and analyzed excitation number and photon coherence spectra for small to medium sized weakly driven dissipative coupled resonator systems described by the Jaynes-Cummings-Hubbard Hamiltonian. Simulations were at all times performed using the full JCH Hamiltonian without approximations, using model parameters realizable by current state-of-the-art technology. We have also presented analogous spectra for resonator arrays governed by the Bose Hubbard model, drawing attention to the differences in experimentally accessible observables between the two models. These differences arise primarily because of the composite nature of the elementary excitations of the JCH Hamiltonian, leading to readily identifiable unique spectral characteristics. Therefore we conclude that when simulating coupled resonator arrays the full JCH Hamiltonian must be used in calculations to properly account for the underlying physics. 

Generalizations of two bosonic interaction-induced non-equilibrium phenomena were observed in the NESS of CRAs modeled retaining the richer JCH physics, with additional tunability.  The fingerprints of the polaritonic equivalents of `fermionic' photons and photon crystallization exhibit subtle departures from the pure bosonic case, detectable via experimentally accessible spectral quantities. 
		
Finally, this work has shown that the combination of the quantum trajectory method with the TEBD algorithm is an especially powerful tool in finding the NESS of open driven dissipative many-body systems. Such a numerical approach can be readily applied to calculate more complex out of equilibrium properties of coupled resonator arrays where an analytic approach is unfeasible. This might include important open problems such as determining the transport properties of linear coupled resonator arrays in the presence of photon loss, or exploring novel one-dimensional quantum states of light that are robust to experimentally realistic decay processes. 

\appendix
\setcounter{section}{1}
\section*{Appendix}
\label{app:MPS_traj}
\emph{Matrix product state quantum trajectory calculations - }  
To solve for NESS expectation values of larger driven dissipative CRAs we employ a stochastic unraveling of the time dependent quantum master equation. This involves propagating independent random wave-function trajectories through time rather than the (much larger) full density matrix \cite{dalibard1992wave, dum1992monte}.  We evolve $R$ stochastic wave-functions $|\Psi_i (t) \rangle$ under the action of a non-Hermitian Hamiltonian $\hat{H}^{\rm eff}_X = \hat{H}_X - i \frac{\gamma_p}{2} \sum_{j=1}^M \hat{a}^\dag_j \hat{a}_j$. The latter terms induce a decay in the wave-function normalization. In practice, quantum jumps corresponding to photon loss from one of the resonators are applied at times $t_{\rm jump}$ when the norm falls below a randomly chosen number $r \in [0,1]$. The particular resonator $j$ is selected by sampling from the probability distribution $P_j = || \hat{a}_j \Psi_i(t_{\rm jump}) \rangle||^2 / \left ( \sum_k || \hat{a}_k \Psi_i(t_{\rm jump}) \rangle||^2 \right )$. After the jump is applied, the wave-function is re-normalized and evolution continues. 

The NESS density matrix $\rho_{\rm ss}$ is calculated by evolving this ensemble of wave-functions to times $t > t_{\rm trans}$ larger than the timescale over which transient dynamics die out, then averaging over realizations as $\bar{\rho}_{\rm ss} \approx (1 / R) \sum_i |\bar{\Psi}_i \rangle \langle \bar{\Psi}_i |$. Here, the over bars denote an additional average over time steps $t_i > t_{\rm trans}$ in the simulation. This important trick is possible due to the ergodicity of the unraveling of the master equation which means that in the NESS the stochastic wave-functions at each time step must average to the true steady state density matrix $\rho_{\rm ss}$ \cite{mejia2007heat,michel2008transport,daley2009atomic}. Hence the trajectory method is far more efficient at simulating non-equilibrium steady states than the transient dynamics. A single trajectory can yield estimates of steady state expectation values and multiple trajectory realizations give an indication of the statistical errors in calculated quantities. Yet another advantage of using a trajectory method here is that NESS expectation values for homogeneous systems can be further averaged over each site in the system and larger simulations gain considerable accuracy from this fact. 

Time evolving the stochastic wave-functions is itself a nontrivial task, as the dimension of Hilbert space is still prohibitively large for a representation in the bare basis. We employ a matrix product state (MPS) \cite{perez2007matrix} ansatz to compress our description of the wave-functions $|\Psi_i(t) \rangle$, and propagate the state in time to near exact accuracy within this representation using the TEBD algorithm \cite{vidal2003efficient, vidal2004efficient}.

We note that in addition to the drastic reduction in computation time and improvement in accuracy afforded by the time averaging procedure, a relatively modest matrix dimension in the MPS system description is sufficient for accurate results. This is a consequence of long range correlations in the system being constantly broken up by the local incoherent processes. A quantitative analysis of the relationship between correlations in individual trajectory wave-functions, and those in the steady state density matrix, will be presented elsewhere \cite{our_future_paper}. 

\emph{Polariton crystallization calculations - } To obtain the steady states used in Fig.~\ref{fig:polariton_crystallization}, we began time averaging at $g (T_{\rm start}) = 500$. We found that $N_T = 5000$ time-steps $g (\Delta t) = 0.5$, with around 100 trajectories retaining around 40 states in the matrix product state representation, were sufficient to obtain acceptable statistical fluctuations ($\approx 0.1\%$) in the NESS expectation values we are interested in. We found that retaining three or four photons in the bare resonator basis was sufficient for systems in the `atomic' regime ($\Delta \geq 0$), but that more were required to properly evaluate steady state quantities for $\Delta < 0$ (results not shown). 

\section*{References}

\bibliography{mybib_cavities}

\end{document}